\documentclass[a4paper]{jpconf}
\usepackage{graphicx}
\usepackage{url}
\RequirePackage{HEPunits}
\newcommand{\eVs}{\eV\squared{}}
\newcommand{\goesto}{\ensuremath{\rightarrow}}
\newcommand{\SuperK}{Super-Kamiokande}
\newcommand{\neut}[1]{\ensuremath{\nu_{#1}}}
\newcommand{\nue}{\neut{e}}
\newcommand{\numu}{\neut{\mu}}
\newcommand{\neutbar}[1]{\ensuremath{\bar{\nu}_{#1}}}
\newcommand{\numubar}{\neutbar{\mu}}
\newcommand{\deltaCP}[1]{\ensuremath{\delta_{CP}}}
\newcommand{\pod}[1]{{P\O{}D}}
\begin{document}
\title{T2K Results and Future Plans}

\author{Francesca Di Lodovico~\footnote{Partially funded by the European Research Council,\
 Grant agreement no. 207282-T2KQMUL.}, for the T2K Collaboration}

\address{School of Physics and Astronomy, Queen Mary University of London, London E1~4NS, UK}

\ead{f.di.lodovico@qmul.ac.uk}

\begin{abstract}
We present the \numu \goesto \nue\ appearance and the \numu\ disappearance results, 
using a total of $1.43 \times 10^{20}$ protons on target collected with the T2K experiment.
T2K is long baseline neutrino experiment in Japan with detectors located at J-PARC, Tokai,
and at Kamioka in the Gifu Prefecture, situated 295 km away from J-PARC.
The muon neutrino beam is produced and measured at the near detectors at J--PARC 
whilst the neutrino rates after oscillation are measured with the \SuperK\ detector, at Kamioka.
A total of six events pass all the selection criteria for \numu \goesto \nue\ oscillations 
at the far detector \SuperK, leading to $0.03(0.04) < \sin^2 2\theta_{13} < 0.28(0.34)$
for \deltaCP\ = 0 and normal (inverted) hierarchy at 90$\%$ C.L.
The \numu\ disappearance analysis excludes no oscillations at 4.3$\sigma$.
At 90$\%$ C.L., the best fit values are $\sin^2 2\theta_{23} > 0.84$ and
$2.1 \times 10^{-3} < \Delta m^2_{23} (\eVs) < 3.1 \times 10^{-3} $.
Finally, we present an overview of the T2K plans from 2011 onwards.
\end{abstract}

\section{Introduction}
The main goal of the T2K (Tokai-to-Kamioka) long baseline neutrino experiment~\cite{Abe:2011ks}
is to measure the appearance of electron neutrinos from a beam of muon neutrinos, thereby
measuring $\theta_{13}$, the last unknown mixing angle in the lepton 
sector.
The results on the muon into electron neutrino appearance~\cite{Abe:2011sj} and the muon neutrino 
disappearance presented in these proceedings are based on the data
collected in the the first two periods of data taking from January 2010 to March 2011.
This is followed by an overview of the future goals for data collections at T2K.

\section{The T2K Experiment}
The T2K experiment's main components are the beam source, the near detector complex
and the far detector. Details of the T2K experimental setup are
described in Ref.~\cite{Abe:2011ks}. Here we provide a brief review of the main components.
The T2K muon neutrino beam is produced at J-PARC\footnote{Japan Proton Accelerator Research Complex 
jointly constructed and operated by KEK and JAEA.} and detected with \SuperK , 
the far detector, located in Kamioka in the Gifu Prefecture 295 km away from J-PARC. 
T2K adopts the off--axis method \cite{Beavis:1995up} to generate a
narrow--band neutrino beam using the new MW--class proton
synchrotron at J-PARC. The neutrino beam is thus directed at an angle 
of 2.5\degree{} with respect to the baseline connecting the 
proton target and \SuperK , corresponding to a peak energy of about
0.6~\GeV, which maximizes the effect of the neutrino oscillation 
at 295~km and minimizes the background to electron-neutrino
appearance detection.
The angle can be reduced down to $2.0$\degree{}, allowing variation of
the peak neutrino energy, if necessary.

The near detector site
at about 280~m from the production target
houses on--axis and off--axis detectors with respect to the direction of the neutrino beam.
The on--axis detector INGRID (Interactive Neutrino GRID)~\cite{bib:INGRID} is composed of an array of 
iron/scintillator sandwiches and its main goal is to measure the neutrino beam
direction and profile.
INGRID can measure the beam centre with a precision better than 28~cm
corresponding to 1~mrad or a shift of around 2$\%$ in the peak of the neutrino
energy spectrum.
The off--axis detector (ND280), see Ref.~\cite{NUFACT11:McCauley} of this conference proceedings, is a multipurpose 
magnetised detector which measures the muon neutrino flux and energy spectrum and intrinsic
electron neutrino contamination in the direction of \SuperK . It also measures the
rates for exclusive neutrino reactions, relevant
to the disappearance and appearance searches at \SuperK .
These measurements are essential
in order to characterize signals and backgrounds that are observed in the
\SuperK\ far--detector.
The off--axis detector measures both charged and neutral particles. 
It consists of a water-scintillator detector optimized to identify the $\pi^0$'s (the \pod{}); a tracker
system consisting of time projection chambers (TPCs)~\cite{bib:TPC} and 
fine grained detectors (FGDs) optimized to study charged current interactions; 
and finally an electromagnetic calorimeter (ECal) that surrounds the \pod{} and the tracker. 
The whole off--axis detector is placed in 
a 0.2~T magnetic field provided by the recycled UA1 magnet, which also serves 
as part of a side muon range detector (SMRD).  

The far detector, \SuperK , is located in the Mozumi mine, near the village of Higashi-Mozumi, 
Gifu, Japan. A detailed description of the detector can be found elsewhere~\cite{fukuda:2002uc}.
Here we provide a summary. The detector cavity lies under the peak of Mt. Ikenoyama, with 1000~m of rock
equivalent to 2700 meters--water--equivalent (m.w.e.) which reduces the
cosmic ray background by around five orders of magnitude compared to at the Earth's surface.

\SuperK\ is a water Cherenkov detector with a fiducial volume (FV) of 22.5~kton within 
its cylindrical inner detector (ID). Surrounding the ID is the 2 m-wide outer detector (OD).
The inner detector (ID) is a cylindrical space
33.8 m in diameter and 36.2 m in height
which currently houses along its inner walls
11,129 inward-facing 50 cm diameter PMTs.
The OD contains along its inner walls 1,885 outward-facing 20 cm diameter PMTs.
The ID and OD boundaries are defined by a cylindrical structure about 50 cm wide.
The front-end readout electronics allows for a zero--deadtime software trigger.
Spill timing information, synchronized by the Global Positioning System (GPS) 
with $<150$~ns precision, is transferred online to \SuperK\ and triggers the recording of
photomultiplier hits within $\pm$500 $\mu$s of the expected arrival time of the
neutrinos.
\SuperK\ has been running since 1996 and has had four distinct running periods. 
The latest period, SK-IV, is running stably and features upgraded PMT readout electronics. 

Construction of the neutrino beamline started in April 2004. 
The complete chain of accelerator and neutrino beamline was successfully commissioned during 2009, and 
T2K began  accumulating neutrino beam data for physics analysis in January 2010.

Construction of the majority of the ND280 detectors was completed in 2009 with the full
installation of INGRID, the central ND280 off-axis sub-detectors (\pod{}, FGD, TPC and downstream ECal) and the SMRD. 
The ND280 detectors began stable operation in February 2010.
The rest of the ND280 detector (the ECals) was completed in the fall of 2010.

The results presented in these proceedings are based on the first two 
physics runs: run~1 (Jan--Jun 2010) and run~2 (Nov 2010--Mar 2011) whereas the 
proton beam power was continually increased and reached
145~kW with $9\times 10^{13}$ protons per pulse by March 2011.
A total of $1.43\times10^{20}$ protons on target (p.o.t.) were collected for analysis,
corresponding to about 2\%\ of the final approved goal for T2K.

\section{Indication of Electron Neutrino Appearance}
\label{sec:appearance}
The \numu \goesto \nue\ analysis at \SuperK\ requests only a single electron-like ($e$-like) ring,
thus selecting a sample enhanced in \nue\ charged-current quasi-elastic interactions (CCQE).
The selection criteria for this analysis were fixed from Monte Carlo (MC) studies before analysing
the data and were optimized for the initial running conditions.

The main backgrounds to the \numu \goesto \nue\ oscillated events are due to the
intrinsic \nue\ contamination in the beam and to the neutral current (NC) interactions 
with a misidentified $\pi^0$.

The observed number of events is compared to expectations based on 
neutrino flux and cross-section predictions for signal and all sources of backgrounds, which are
corrected using an inclusive \numu\ charged-current (CC) measurement in the off--axis near detector.

We compute the neutrino beam fluxes starting from models and tune them to experimental data.
A detailed description can be found in Ref.~\cite{NUFACT11:Galymov} of this conference proceedings.
The estimated uncertainties of the intrinsic \numu\ and \nue\ fluxes below 1~\GeV\
are around 14\%.
Above 1~\GeV, the intrinsic \nue\ flux error is dominated by the uncertainty on the kaon
production rate with resulting errors of 20--50\%.

\begin{figure}[!tbp]
  \centering
  \includegraphics[width=22pc]{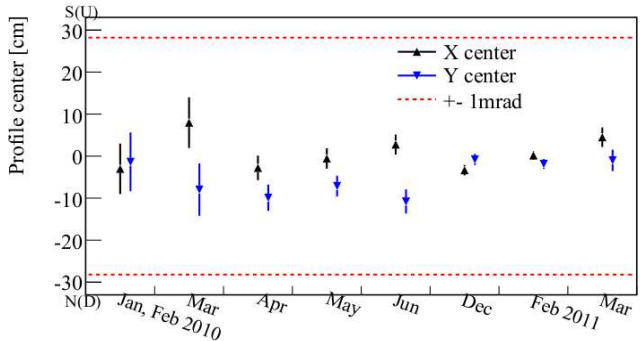}\hspace{2pc}%
  \begin{minipage}[b]{12pc}\caption{\label{fig:ingrideventsteer}Beam centering stability in horizontal (x, South--North) and vertical (y, Down--Up) directions as a function of time, as measured by INGRID. Errors shown are only statistical.
  The dashed lines correspond to a change of beam direction by $\pm$1~mrad.}
\end{minipage}
\end{figure}

The neutrino  beam profile and its absolute rate ($1.5$~events$/10^{14}$~p.o.t.) as measured by INGRID
were stable and consistent with expectations. 
The beam profile center (Fig.~\ref{fig:ingrideventsteer}) indicates that beam steering was better than $\pm 1$~mrad. 
The correlated systematic error is $\pm0.33(0.37)$~mrad for the horizontal(vertical) direction.
The error on the \SuperK\ position relative to the beamline elements was obtained
from a dedicated GPS survey and is negligible.

The NEUT MC event generator~\cite{hayato:neut}, which has been
tuned with recent neutrino interaction data in an energy region
compatible with T2K, is used to simulate neutrino interactions in the near and far
detectors.
The GENIE~\cite{Andreopoulos:2009rq}
generator provides a separate cross-check of the assumed cross-sections and
uncertainties, and yields consistent results.  

A list of reactions and their uncertainties relative to the CCQE total
cross-section is shown in Table~\ref{tab:systneut}.   An energy--dependent error
on CCQE is assigned to account for the uncertainty in the low
energy cross--section, especially for the different
target materials between the near and far detectors.

 \begin{table}[!tbp]
   \centering
   \caption{Summary of systematic uncertainties for the relative rate of
       different charged-current (CC) and neutral-current (NC) reactions
       to the rate for CCQE.}
   \begin{tabular}{lc}
     \hline
     \hline
   Process & Systematic error \\
       \hline
       CCQE & energy-dependent (7\% at 500 \MeV) \\
       CC 1$\pi$ & 30\% ($E_\nu<2$~\GeV) -- 20\% ($E_\nu>2$~\GeV) \\
       CC coherent $\pi^\pm$ & 100\%  \\
       CC other & 30\% ($E_\nu<2$~\GeV) -- 25\% ($E_\nu>2$~\GeV) \\
       NC 1$\pi^0$ & 30\% ($E_\nu<1$~\GeV) -- 20\% ($E_\nu>1$~\GeV) \\
       NC coherent $\pi$ & 30\% \\
       NC other $\pi$ & 30\% \\
       FSI & energy-dependent (10\% at 500 \MeV) \\
         \hline
     \hline
   \end{tabular}
   \label{tab:systneut}
 \end{table}

An inclusive $\nu_\mu$~CC measurement in the off-axis near detector
is used to constrain the expected event rate at the far detector.
The results are described in Ref.~\cite{NUFACT11:Roberge} of this conference proceedings.
The measured data/MC ratio is
\begin{equation}
R^{\mu,Data}_{ND}/R^{\mu,MC}_{ND} = 1.036 \pm 0.028 (\mathrm{stat.}) ^{+0.044}_{-0.037} (\mathrm{det.syst.})\pm0.038(\mathrm{phys.syst.}),
\label{eq:ratiodmc}
\end{equation}
where $R_{ND}^{\mu,Data}$ and $R_{ND}^{\mu,MC}$ are the p.o.t. normalized rates of $\nu_\mu$~CC interactions in data and MC.
The detector systematic errors mainly come from tracking and particle identification efficiencies,
and physics uncertainties are related to the interaction modeling.
Uncertainties that effectively cancel between near and far detectors were omitted.

\begin{figure}[!tbp]
\begin{minipage}{15pc}
\includegraphics[width=15pc]{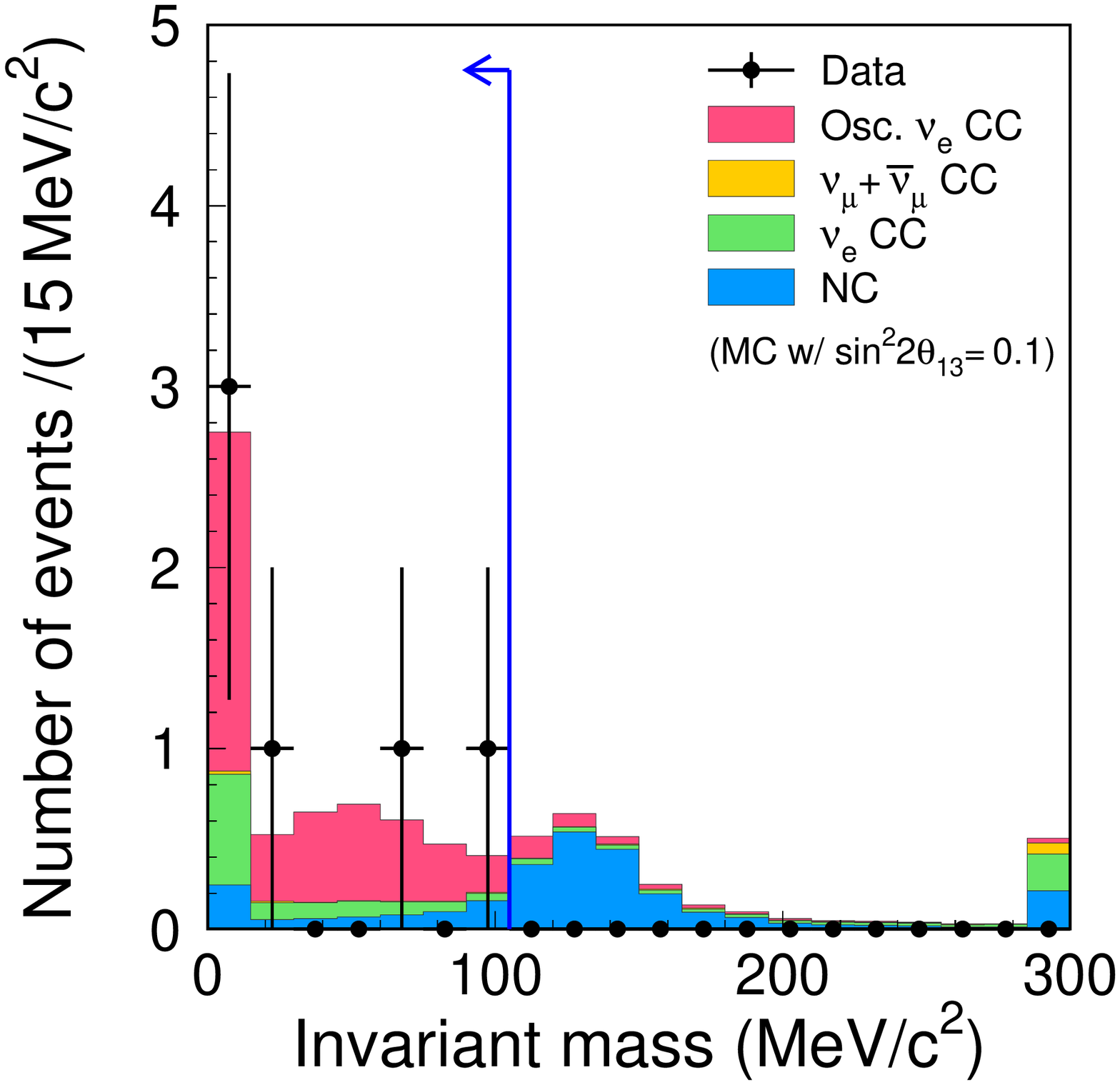}
\caption{\label{fig:polfit}Invariant mass distribution reconstructing each event as two rings.
   The data are shown using points with statistical error bars and the MC predictions are the histograms.
   The last bin shows overflow entries and the vertical line the applied cut at 105 \MeV.
   }
\end{minipage}\hspace{4pc}%
\begin{minipage}{13pc}
\includegraphics[width=15pc]{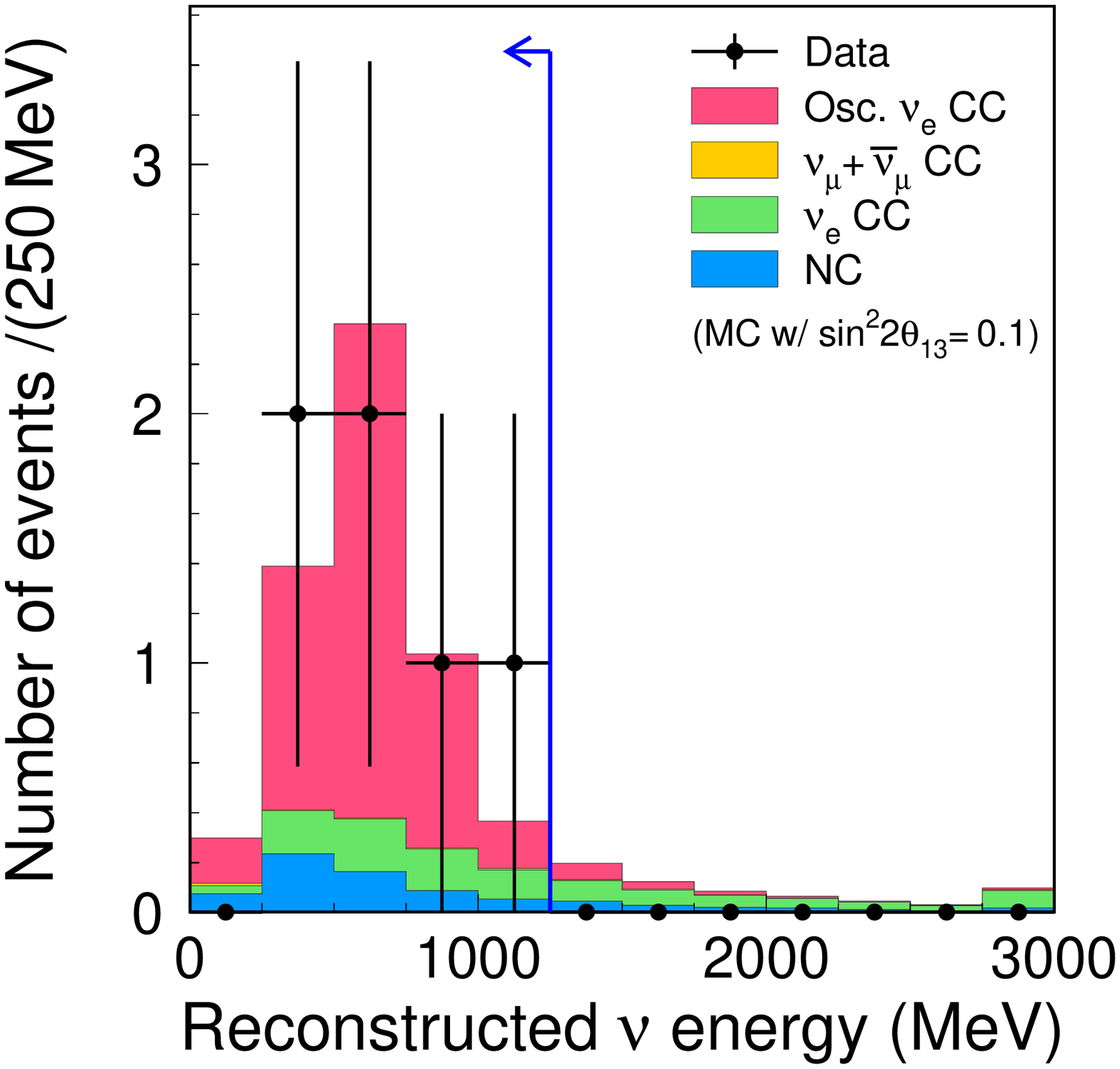}
\caption{\label{fig:recoene}Same as Fig.~\ref{fig:polfit} for the
  reconstructed neutrino energy spectrum of the events
  which pass all \nue\ appearance signal selection criteria with the 
  exception of the energy cut.
  The vertical line shows the applied cut at 1250 \MeV.
  }
\end{minipage} 
\end{figure}

At the far detector, we extract a fully-contained fiducial volume (FCFV) sample by requiring
no event activity in either the OD or 
in the 100 $\mu$s  before the event trigger time,
at least $30$ \MeV\ electron-equivalent energy deposited in the ID (defined as visible energy $E_{vis}$), 
and the reconstructed vertex in the fiducial region.
The data have 88 such FCFV events that are within the timing range 
from $-$2 to 10~$\mu$s around the beam trigger time. 
The accidental contamination from non--beam related events
is determined from the sidebands to be $0.003$~events.
A Kolmogorov-Smirnov (KS) test of the observed number of FCFV events as a function of accumulated 
p.o.t. is compatible with the normalized event rate being constant ($p$-value = 0.32).
Forty-one events are reconstructed with a single ring, and eight of those are $e$-like.
Six of these events have $E_{vis}>100$~\MeV\ and no delayed-electron signal.
To suppress misidentified $\pi^0$ mesons, the reconstruction of two
rings is forced by comparison of the observed and expected light patterns calculated 
under the assumption of two showers, and a cut on the two-ring invariant mass 
$M_{inv}<105$~\MeV$/c^2$ is imposed. No events are rejected (Fig.~\ref{fig:polfit}).
Finally, the neutrino energy $E^{rec}_{\nu}$ is computed using the reconstructed
momentum and direction of the ring, by assuming quasi--elastic kinematics
and neglecting Fermi motion.
No events are rejected by requiring $E^{rec}_{\nu}<1250$~\MeV, whose purpose
was to suppress events from the intrinsic $\nu_e$ component arising
primarily from kaon decays (Fig.~\ref{fig:recoene}).
The data and MC reductions after each selection criterion are shown in Table~\ref{table:eventsummary}.
The $\nu_e$ appearance signal efficiency is estimated from MC to be 66\% while
rejection for $\nu_\mu+\bar\nu_\mu$ CC, intrinsic
\nue CC, and NC are $>99$\%, 77\%, and 99\%, respectively. 
Of the surviving background NC interactions constitute 46\%,
of which 74\% are due to $\pi^0$ mesons
and 6\% originate from single gamma production.
 \begin{table}[!tbp]
  \centering
  \caption{Event reduction for the \nue\ appearance search at the \SuperK .
  The numbers of data and $\nu_\mu$~CC, intrinsic \nue CC, 
 NC, and \nue\ CC signal MC expected events are presented after each selection criterion.
 All MC CC samples include three-flavor oscillations for $\sin^2 2\theta_{13}$ = 0.1 and \deltaCP\ = 0.
 }
  \begin{tabular}{lccccc}
   \hline
   \hline
                        & Data & \numu{}CC & \nue{}CC & NC &\numu$\rightarrow$\nue{}CC \\
   \hline
   (0)  interaction in FV   &  n/a    &       67.2    &      3.1   &    71.0    &       6.2 \\
   (1)~fully-contained FV & 88 & 52.4 & 2.9 & 18.3 & 6.0 \\
   (2)~single ring          & 41 & 30.8 & 1.8 & 5.7 & 5.2 \\
   (3)~$e$-like        &  8 &  1.0 & 1.8 & 3.7 & 5.2 \\
   (4)~$E_{vis}>100$~\MeV   &  7 &  0.7 & 1.8 & 3.2 & 5.1 \\
   (5)~no delayed electron    &  6 &  0.1 & 1.5 & 2.8 & 4.6 \\
   (6)~non-$\pi^0$-like     &  6 &  0.04 & 1.1 & 0.8 & 4.2 \\
   (7)~$E^{rec}_{\nu}<1250$~\MeV        &  6 &  0.03 & 0.7 & 0.6 & 4.1 \\
   \hline
   \hline
  \end{tabular}
  \label{table:eventsummary}
 \end{table}

Examination of the six data events shows properties consistent 
with $\nu_{e}$ CC interactions. 
The distribution of the cosine of the opening angle between the ring and the incoming beam direction
is consistent with CCQE events.
The event vertices in cylindrical coordinates ($R$,$\phi$,$z$)
show that these events are clustered at large $R$, 
near the edge of the FV in the upstream beam direction.
A KS test on the $R^2$ distribution of our final events yields a $p$-value of 0.03.
If this was related to contamination from penetrating particles produced in upstream neutrino
interactions,  then the ID region outside the FV should show evidence for such events, 
however this is not observed.
In addition, an analysis of the neutrino interactions occurring in the OD volume is consistent with expectations.

To compute the expected number of events at the far detector $N_{SK}^{exp}$,
we use the near detector $\nu_\mu$~CC interaction rate measurement as normalization, and 
the ratio of expected events in the near and far detectors, where common systematic errors cancel.
Using Eq.~\ref{eq:ratiodmc}, this can be expressed as:
\begin{equation}
N_{SK}^{exp} =\left({R_{ND}^{\mu,Data}}/{R_{ND}^{\mu,MC}} \right) \cdot N_{SK}^{MC},
\label{eq:nskexp}
\end{equation}
where $N_{SK}^{MC}$ is the MC number of events expected in the far detector.
Due to the correlation of systematic errors in the near and far detector samples, Eq.~\ref{eq:nskexp}
reduces the uncertainty on the expected number of events. 
Event rates are computed 
incorporating three-flavor oscillation probabilities and matter effects~\cite{PhysRevD.22.2718}
with $\Delta m^2_{12}$ = $7.6\times 10^{-5}$~eV$^2$,
$\Delta m^2_{23}$ = $+2.4\times 10^{-3}$~eV$^2$, $\sin^2 2\theta_{12}$ = 0.8704,
$\sin^2 2\theta_{23}$ = 1.0, an average Earth
density $\rho$ = 3.2~g/cm$^3$ and \deltaCP\ = 0 unless otherwise noted.
The expectations  are
0.03(0.03)  \numu + \numubar CC, 0.8(0.7) intrinsic \nue CC, 
and 0.1(4.1) \numu $\rightarrow$ \nue oscillation events for $\sin^2 2\theta_{13}$ = 0(0.1),
and 0.6~NC events.
As shown in Table~\ref{table:syserr}, 
the total systematic uncertainty on $N_{SK}^{exp}$ depends on $\theta_{13}$.
Combining the above uncertainties the total
far detector systematic error contribution to $\delta N_{SK}^{exp}/N_{SK}^{exp}$ is 14.7\%(9.4\%)
for $\sin^22\theta_{13}$ = 0(0.1). 

\begin{table}[btp]
\begin{center}
\caption{\small Contributions from various sources
 and the total relative uncertainty for $\sin^22\theta_{13}$ = 0 and 0.1, and \deltaCP\ = 0.}
\begin{tabular}{lrrr}
\hline \hline
         Source        & & ~~~$\sin^22\theta_{13}$ = 0  & ~~~$\sin^22\theta_{13}$ = 0.1      \\
\hline
(1)~neutrino flux        & &  $\pm$  8.5\%   & $\pm$ 8.5\%  \\
(2)~near detector    & &  ${}^{+5.6}_{-5.2}$\%  &  ${}^{+5.6}_{-5.2}$\%   \\
(3)~near det. statistics    & & $\pm$ 2.7\%    & $\pm$ 2.7\%  \\
(4)~cross section    & &   $\pm$ 14.0\%   & $\pm$ 10.5\%  \\
(5)~far detector     & &  $\pm$ 14.7\%    & $\pm$ 9.4\%  \\
\hline
Total  $\delta N_{SK}^{exp}/N_{SK}^{exp}$         & &  ${}^{+22.8}_{-22.7}$\%   &  ${}^{+17.6}_{-17.5}$\%  \\
\hline \hline
\end{tabular}
\label{table:syserr}
\end{center}
\end{table}

\begin{figure}[h]
\begin{tabular}{cc}
\includegraphics[width=16pc]{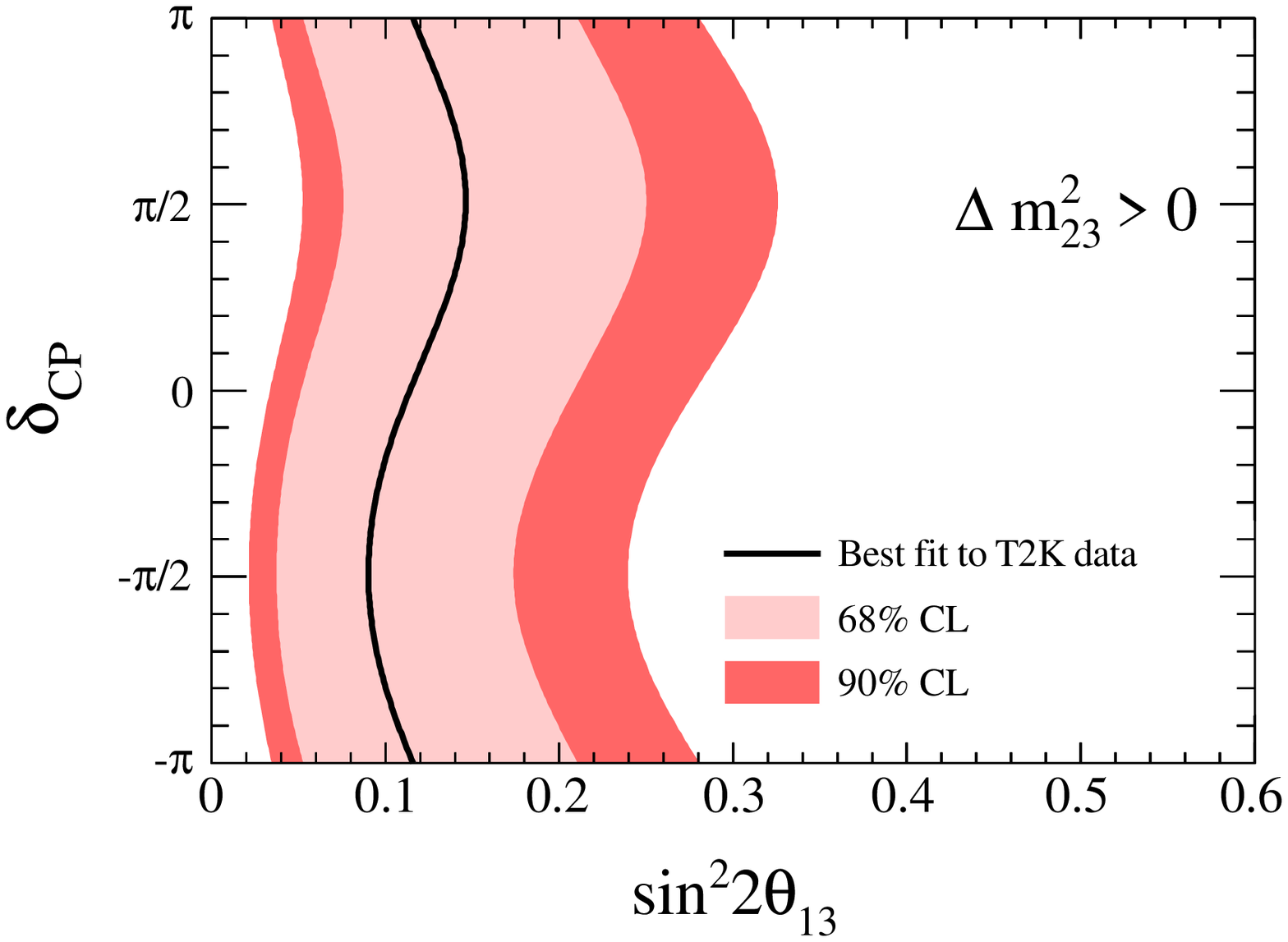}\hspace{2pc}% &
\includegraphics[width=16pc]{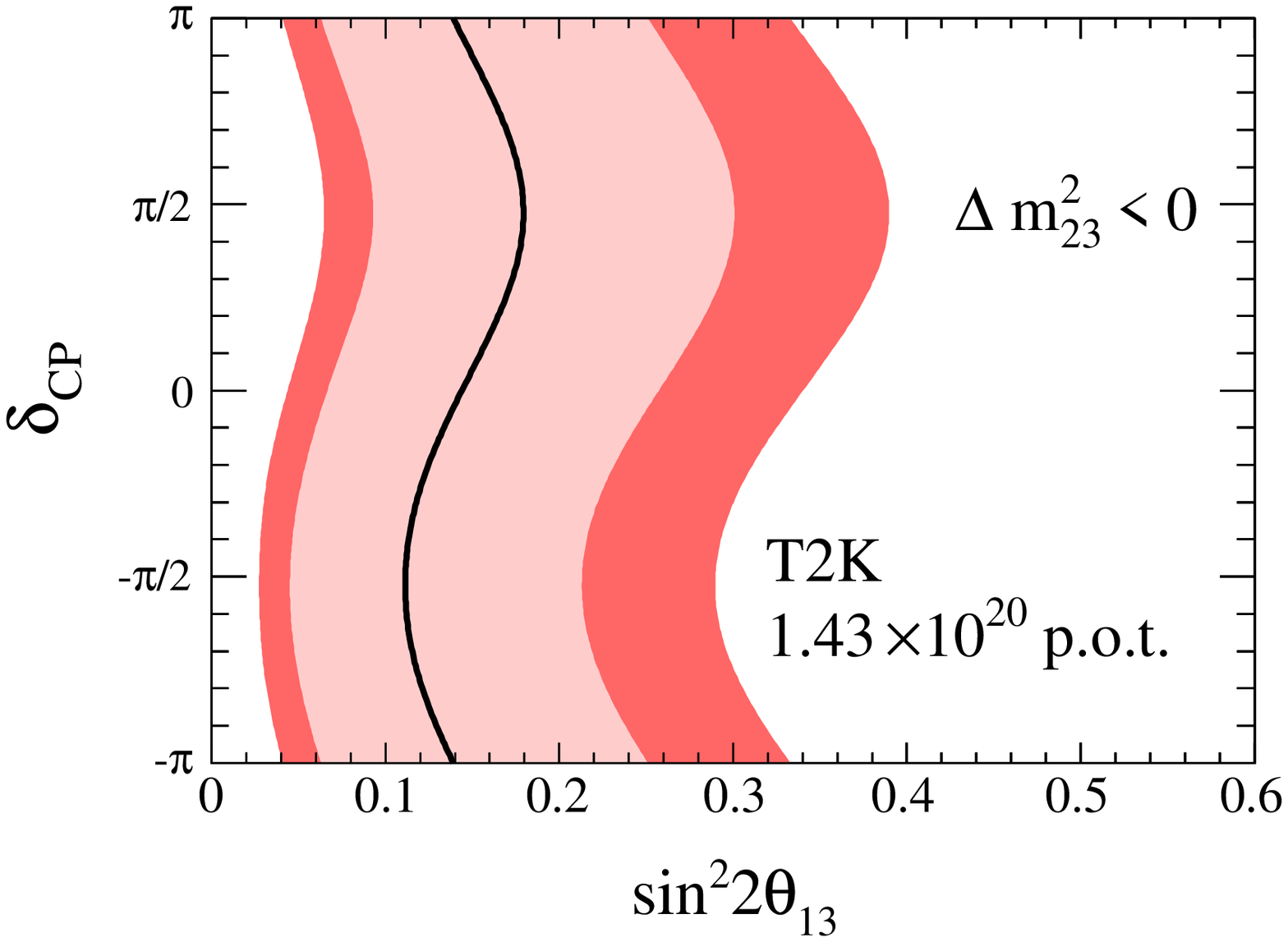}\hspace{2pc}%\\
\end{tabular}
\caption{\label{fig:normcontour}The 68\% and 90\%~C.L. regions for $\sin^{2}2\theta_{13}$ for each value of \deltaCP , consistent with the observed number of events in the three-flavor oscillation case for normal (left) and inverted (right) mass hierarchy. The other oscillation parameters are fixed (see text). The best fit values are shown with solid lines.}
\end{figure}

Our oscillation result is based entirely on comparing the number of $\nu_{e}$ candidate events with predictions,
varying $\sin^22\theta_{13}$ for each \deltaCP\ value.
Including systematic uncertainties the expectation is 1.5$\pm$0.3(5.5$\pm$1.0)
events for $\sin^2 2\theta_{13}$ = 0(0.1).
The probability to observe six or more candidate events is 7$\times$10$^{-3}$ for $\sin^2 2\theta_{13}$ = 0.
Thus, we conclude that our data indicate $\nu_e$ appearance from a $\nu_\mu$ neutrino beam. 
At each oscillation parameter point, a probability distribution for the expected number of events is 
constructed, incorporating systematic errors~\cite{cite:conradetal}, which is used to 
make the confidence interval (Fig.~\ref{fig:normcontour}), following the unified ordering prescription of 
Feldman and Cousins~\cite{cite:feldman_cousins}.

This result converted into a confidence interval yields $0.03(0.04)<\sin^2 2\theta_{13}$ $<$ 0.28(0.34) at 90\%~C.L. for
$\sin^22\theta_{23}$ = 1.0,
$|\Delta^2 m_{23}^2|$ = 2.4$\times$10$^{-3}$ eV$^2$, \deltaCP\ = 0 and
for normal (inverted) neutrino mass hierarchy. Under the same assumptions, the best fit points are 0.11(0.14), respectively.
For non--maximal $\sin^22\theta_{23}$, the confidence intervals remain unchanged to first order by replacing
$\sin^22\theta_{13}$ by $2\sin^2\theta_{23}\sin^22\theta_{13}$.
More data are required to firmly establish $\nu_e$ appearance and to better determine the angle $\theta_{13}$.
These results have been published in Ref.~\cite{Abe:2011sj} and are consistent with the results published subsequently by the MINOS collaboration~\cite{Adamson:2011qu} that measures $2\sin^2\theta_{23}\sin^22\theta_{13} < 0.12(0.20)$ at 90\%\ C.L. for \deltaCP\ = 0 and the normal (inverted) mass hierarchy. More details are given in Ref.~\cite{NUFACT11:MINOS} in this conference proceedings.

\section{Muon Neutrino Disappearance}
We present preliminary results on the \numu\ disappearance analysis in the following. 
The \numu\ analysis follows the same steps of the \numu \goesto \nue\ appearance analysis up to the single ring selection criteria, that corresponds to the event reduction criteria in Table~\ref{table:eventsummary} number (2).
From the selected events, that are predominantly CCQE, thirty--three \numu --like events are indentified. 
By adding a further cut on the minimum muon momentum and on the observed delayed electrons, the CCQE purity is increased with a negligible decrease in efficiency. A total of thirty--one single $\mu$--ring events are observed with an expected number of events of about 104 without oscillations.

The same strategy as for the \numu \goesto \nue\ analysis with regard to the beam simulation, the neutrino interactions and the treatment of the ND280 results is adopted (see Section~\ref{sec:appearance}).
The data and MC reductions after each selection criterion are shown in Table~\ref{table:eventnumusummary}.
%The rejection for the intrinsic \nue\ CC and NC backgrounds are $>99$\%\ and 98\%, respectively. 
%
 \begin{table}[!tbp]
  \centering
  \caption{Event reduction for the \numu\ disappearance search at \SuperK . The selection criteria (0) to (2) are shown in Table~\ref{table:eventsummary}. 
   The numbers of data and
 $\nu_\mu$~CCQE, \numu{}CC non--CCQE,  
 intrinsic \nue CC and NC MC expected events are presented after each selection criterion.
 All MC CC samples include three-flavor oscillations for $\sin^2 2\theta_{23}$ = 1 and $\Delta m^2_{23}$ = 0.0024 \eVs .
 }
  \begin{tabular}{lccccc}
   \hline
   \hline
                               & Data & \numu{}CCQE & \numu{}CC non--CCQE & \nue CC & NC \\
   \hline
   (3)~$\mu$-like              &  33 &  17.6 & 12.4 & $<$0.1 & 1.9 \\
   (4)~$P_{\mu}>200$~\MeV       &  33 &  17.5 & 12.4 & $<$0.1 & 1.9 \\
   (5)~0 or 1 delayed electron &  31 &  17.3 & 9.2  & $<$0.1 & 1.8 \\
   \hline
   \hline
  \end{tabular}
  \label{table:eventnumusummary}
 \end{table}

\begin{figure}[!tbp]
\begin{minipage}{16.pc}
\includegraphics[width=16pc]{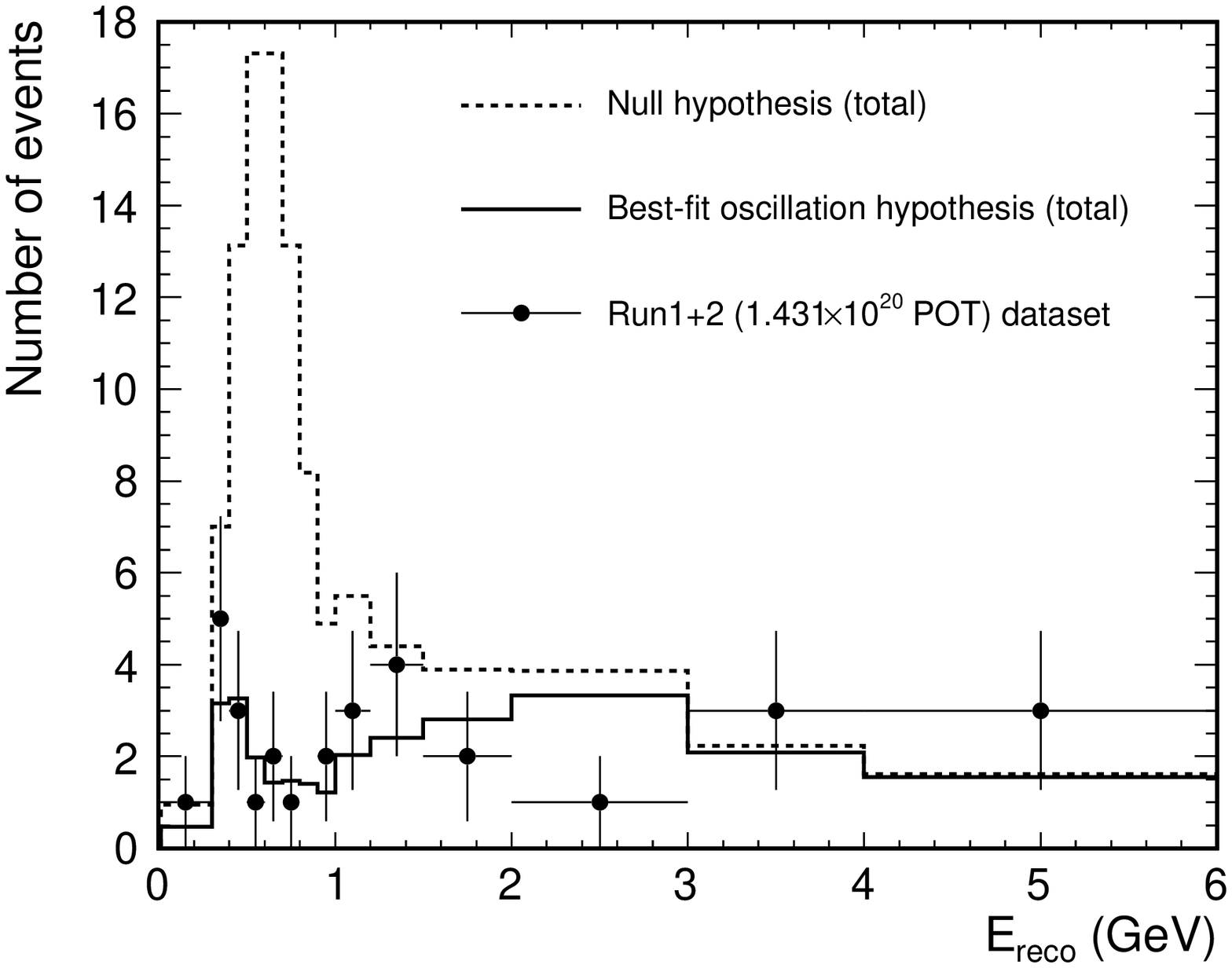}
\caption{\label{fig:muenergy}
Reconstructed neutrino energy spectrum of the thirthy--one single $\mu$--like ring events (points with statistical error bars), with no oscillations and best--fit MC prediction (histograms). 
   }
\end{minipage}\hspace{2pc}%
\begin{minipage}{16pc}
\includegraphics[width=16.5pc]{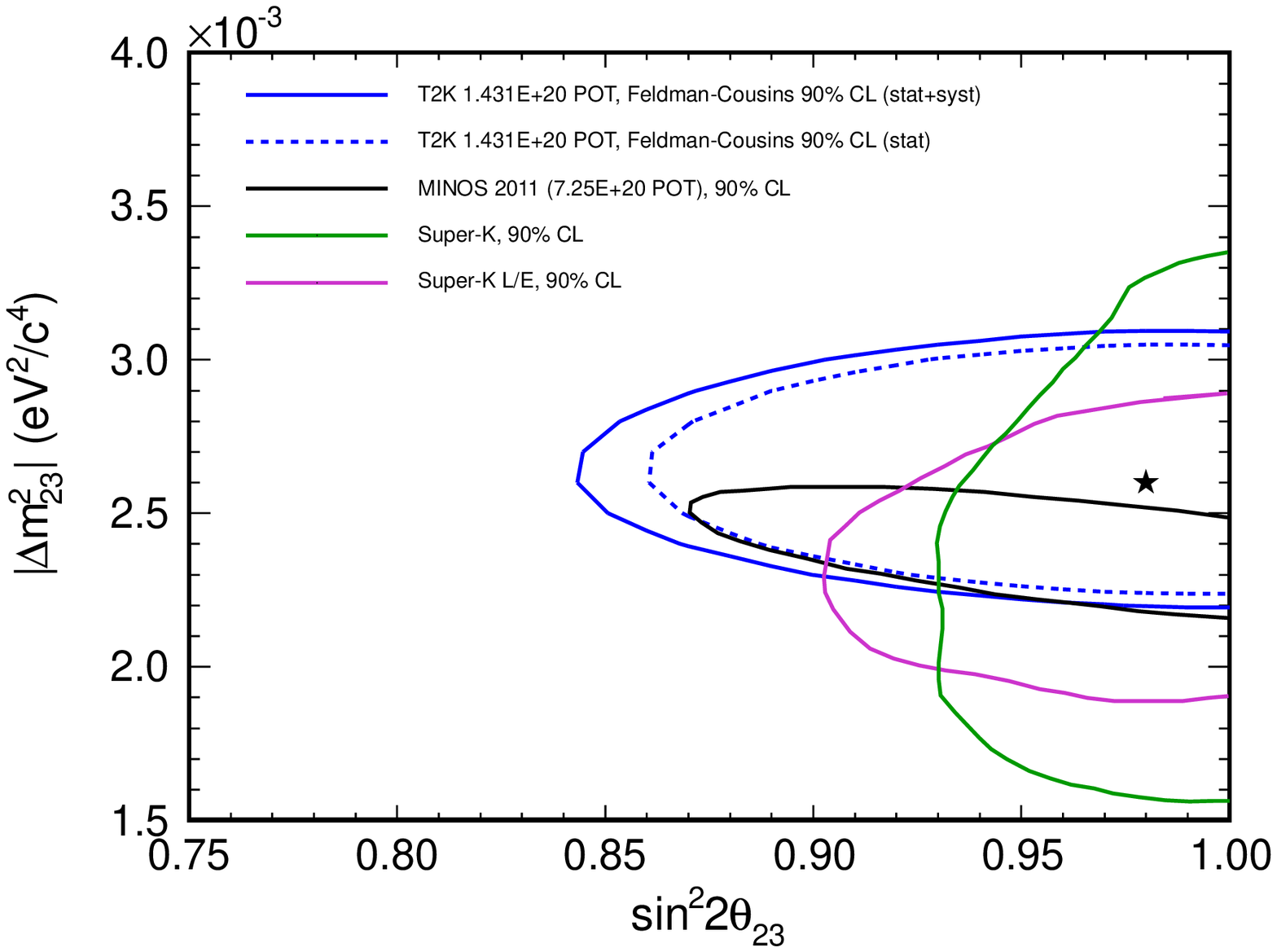}
\caption{\label{fig:mucontour}The 90\%\ \numu --disappearance confidence regions (with and without systematics) compared with \SuperK\ and MINOS results. The star indicates the best fit point. }
\end{minipage} 
\end{figure}

The reconstructed energy spectrum of the thirty--one single $\mu$--like ring events in the combined run 1 and 2 dataset, with prediction for no oscillations, and best--fit prediction is presented in Figure~\ref{fig:muenergy}. 
The statistical significance of the observed deficit without oscillations is 4.3$\sigma$.
The energy spectrum in Figure~\ref{fig:muenergy} is shown in a variable binning scheme for the purpose of illustration only; the actual analysis used equal--sized 50 \MeV\ bins. The best fit parameters are $\sin^22\theta_{23}$ = 0.98, $|\Delta m^2_{23}|$ = 2.6$\times$10$^{-3}$ \eVs , and the corresponding 90\%\ Feldman-Cousins confidence regions (with and without systematic errors) for the 2--flavour \numu\ --disappearance fit are shown in Figure~\ref{fig:mucontour}. We also perform an independent oscillation analysis that gives very similar results: $\sin^22\theta_{23}$ = 0.99, $|\Delta m^2_{23}|$ = 2.6$\times$10$^{-3}$ \eVs\ with a statistical significance of the observed deficit without oscillations of 4.4$\sigma$. Our results are consistent with the ones from \SuperK\cite{bib:SuperKatm} and MINOS~\cite{Adamson:2011ig}.

\section{Future Prospects}
The T2K data taking stopped abruptly on March 11 2011 due to the
Great East Japan Earthquake. 
%Assessment of the damage and recovery started as soon as possible thereafter.
No serious damages were found in the accelerator complex, the
neutrino beamline or the near detectors. The Super-Kamiokande detector
was not affected by the earthquake. 
We are on schedule to resume operations at J-PARC in December 2011 with
the T2K experiment expected to start data taking as soon as possible
early 2012.
The current goal is to collect $1 \times 10^{21}$ p.o.t. by summer 2013.
Such a data sample will allow the confirmation of a non-zero value of $\theta_{13}$
with 5 sigma statistical significance at the present best fit value.

\section{Conclusions}
The T2K experiment, which comprises a new neutrino beamline and new 
near detector complex at J-PARC and the upgraded Super-Kamiokande detector, 
has been constructed successfully and collected data from January 2010 to March 2011
for a total of $1.43\times10^{20}$ p.o.t.
We published our first results on the \nue\ appearance analysis~\cite{Abe:2011sj} where we report the observation of six single ring $e$-like events. The probability to observe six or more events is 7$\times$10$^{-3}$ for $\sin^2 2\theta_{13}$ = 0.
This result yields $0.03(0.04)<\sin^2 2\theta_{13}$ $<$ 0.28(0.34) 
at 90\%~C.L. for $\sin^22\theta_{23}$ = 1.0, $|\Delta m^2_{23}|$ = 2.4$\times$10$^{-3}$ \eVs , \deltaCP\ = 0 and
for normal (inverted) neutrino mass hierarchy. Under the same assumptions, the best fit points are 0.11(0.14), respectively.
%More data are required to firmly establish $\nu_e$ appearance and to better determine the angle $\theta_{13}$.
These results are consistent with the results published subsequently by MINOS.

We presented preliminary results on the \numu\ disappearance analysis using the same data set.
A total of thirty-one single ring $\mu$-like events are observed at the far detector, yielding to 
the best fit parameters of $\sin^22\theta_{23}$ = 0.98, $|\Delta m^2_{23}|$ = 2.6$\times$10$^{-3}$ \eVs , consistent with the results from \SuperK\ and MINOS.

T2K has collected 2\%\ of its projected data and is on schedule to restart data taking early 2012. J-PARC will start operation in December 2011.


\section*{References}
\begin{thebibliography}{9}
%\cite{Abe:2011ks}
\bibitem{Abe:2011ks}
  K.~Abe {\it et al.} [ T2K Collaboration ],
  %``The T2K Experiment,''
  accepted for publication in Nucl. Instrum. Methods, article in press,
  [arXiv:1106.1238 [physics.ins-det]].

\bibitem{Abe:2011sj}
  K.~Abe {\it et al.} [ T2K Collaboration ],
  %``Indication of Electron Neutrino Appearance from an Accelerator-produced Off-axis Muon Neutrino Beam,''
  Phys.\ Rev.\ Lett.\  {\bf 107 } (2011)  041801.
  [arXiv:1106.2822 [hep-ex]].

\bibitem{Beavis:1995up}
D.~Beavis, A.~Caroll, I.~Chiang {\it et al.} [ E889 Collaboration ],
 %''Long Baseline Neutrino Oscillation Experiment at the AGS (Proposal E889)''
Physics Design Report {\bf BNL 52459} (1995).

\bibitem{bib:INGRID}
M.~Otani {\it et al.},
%''Design and construction of INGRID neutrino beam monitor for T2K neutrino experiment ''
Nuclear Instruments and Methods in Physics Research Section A {\bf 623}, 368 (2010).  

\bibitem{NUFACT11:McCauley}
N.~McCauley,
``Performance of the T2K Near Detectors'',
this conference, XIIIth International Workshop on Neutrino Factories, Super beams and Beta beams, proceedings.

\bibitem{bib:TPC}
N.~Abgrall {\it et al.}, 
Nuclear Inst. and Methods in Physics Research Section A {\bf 637}, 25 (2011).

\bibitem{fukuda:2002uc}
Y.~Fukuda {\it et al.}  [ Super-Kamiokande Collaboration ],
Nucl. Instrum. Meth. {\bf A501}, 418 (2003).
%
%\bibitem{Itow:2001ee}
%Y.~Itow {\it et al.}  [ Super-Kamiokande Collaboration ], The JHF-Kamioka neutrino
%project, 2001, ArXiv hep-ex/0106019.

\bibitem{NUFACT11:Galymov}
V.~Galymov,
``Predicting Neutrino Flux for the T2K Experiment'',
this conference, XIIIth International Workshop on Neutrino Factories, Super beams and Beta beams, proceedings.

\bibitem{hayato:neut}
Y.~Hayato, Nucl.Phys.(Proc. Suppl.) {\bf B112}, 171 (2002).

\bibitem{Andreopoulos:2009rq}
  C.~Andreopoulos, A.~Bell, D.~Bhattacharya, F.~Cavanna, 
J.~Dobson {\it et al.}, Nucl.Instrum.Meth. {\bf A614}, 87 (2010), arXiv:0905.2517 [hep-ph].

\bibitem{NUFACT11:Roberge}
D.G.~Brook-Roberge,
``Neutrino Interaction Measurements Using the T2K Near Detector'',
this conference, XIIIth International Workshop on Neutrino Factories, Super beams and Beta beams, proceedings.

\bibitem{PhysRevD.22.2718}
V.~Barger, K.~Whisnant, S.~Pakvasa, and R.J.N.~Phillips, Phys. Rev. {\bf D22}, 2718 (1980).

\bibitem{cite:conradetal}
J.~Conrad, O.~Botner, A.~Hallgren, and C.~Perez de los Heros, Phys. Rev. {\bf D67}, 012002 (2003),                   
arXiv:hep-ex/0202013 [hep-ex].                                                                   

\bibitem{cite:feldman_cousins}
G.J.~Feldman and R.C.~Cousins, Phys. Rev. {\bf D57}, 3873 (1998),                   
arXiv:hep-ex/0202013 [hep-ex].                                                                   

%\cite{Adamson:2011qu}
\bibitem{Adamson:2011qu}
  P.~Adamson {\it et al.} [ MINOS Collaboration ],
  %``Improved search for muon-neutrino to electron-neutrino oscillations in MINOS,''
  %Submitted to: Phys.Rev.Lett..
  [arXiv:1108.0015 [hep-ex]].

\bibitem{NUFACT11:MINOS}
J.~Hartnell,
``MINOS, NOVA, MINOS+'', and J.~Nelson, ``Improved search for muon-neutrino to electron-neutrino oscillations in MINOS'', this conference, XIIIth International Workshop on Neutrino Factories, Super beams and Beta beams, proceedings.


\bibitem{bib:SuperKatm}
Y.~Ashie {\it et al.}  [ Super-Kamiokande Collaboration ], Phys. Rev. Lett. {\bf 93}, 101801 (2004);
Y.~Ashie {\it et al.}  [ Super-Kamiokande Collaboration ], Phys. Rev. {\bf D71}, 112005 (2005).
J. Hosaka {\it et al.}  [ Super-Kamiokande Collaboration ], Phys. Rev. {\bf D74}, 032002 (2006).

%\cite{Adamson:2011ig}
\bibitem{Adamson:2011ig}
  P.~Adamson {\it et al.} [ The MINOS Collaboration ],
  %``Measurement of the neutrino mass splitting and flavor mixing by MINOS,''
  Phys.\ Rev.\ Lett.\  {\bf 106 } (2011)  181801,
  [arXiv:1103.0340 [hep-ex]].

\end{thebibliography}
\end{document}